\documentclass[AMA,STIX2COL]{MRM}
\articletype{PRE-PRINT -- SUBMITTED TO MAGNETIC RESONANCE IN MEDICINE}%

\usepackage{enumitem}
\usepackage{upgreek}
\usepackage[font=footnotesize,labelfont=bf]{caption}
\graphicspath{{./figures/}}
\usepackage{xr}

\received{}
\revised{}
\accepted{}
\makeatletter
\newcommand*{\addFileDependency}[1]{
\typeout{(#1)}
\@addtofilelist{#1}
\IfFileExists{#1}{}{\typeout{No file #1.}}
}\makeatother

\newcommand*{\myexternaldocument}[1]{%
\externaldocument{#1}%
\addFileDependency{#1.tex}%
\addFileDependency{#1.aux}%
}

\myexternaldocument{./SupportingInformation}

\begin{document}

\title{Bias-Reduced Neural Networks for Parameter Estimation in Quantitative MRI}

\author[1,2,3]{Andrew Mao}{\orcid{0000-0002-1398-0699}}

\author[1,2]{Sebastian Flassbeck}{\orcid{0000-0003-0865-9021}}

\author[1,2]{Jakob Assl\"ander}{\orcid{0000-0003-2288-038X}}

\authormark{Andrew Mao \textsc{et al}}

\address[1]{Bernard and Irene Schwartz Center for Biomedical Imaging, Department of Radiology, New York University Grossman School of Medicine, New York, New York}
\address[2]{Center for Advanced Imaging Innovation and Research (CAI$^2$R), Department of Radiology, New York University Grossman School of Medicine, New York, New York}
\address[3]{Vilcek Institute of Graduate Biomedical Sciences, New York University Grossman School of Medicine, New York, New York}

\corres{Andrew Mao. \email{andrew.mao@nyumc.org}}

\finfo{\fundingAgency{National Institutes of Health} Grant Numbers: \fundingNumber{F30~AG077794}, \fundingNumber{R01~NS131948}, \fundingNumber{T32~GM136573}, \fundingNumber{P41~EB017183}}

\abstract[]{\vspace{-3mm}
\section{Purpose} To develop neural network (NN)-based quantitative MRI parameter estimators with minimal bias and a variance close to the Cram\'er-Rao bound.
\section{Theory and Methods} We generalize the mean squared error loss to control the bias and variance of the NN's estimates, which involves averaging over multiple noise realizations of the same measurements during training. Bias and variance properties of the resulting NNs are studied for two neuroimaging applications.
\section{Results} In simulations, the proposed strategy reduces the estimates' bias throughout parameter space and achieves a variance close to the Cram\'er-Rao bound. In vivo, we observe good concordance between parameter maps estimated with the proposed NNs and traditional estimators, such as non-linear least-squares fitting, while state-of-the-art NNs show larger deviations.
\section{Conclusion} The proposed NNs have greatly reduced bias compared to those trained using the mean squared error and offer significantly improved computational efficiency over traditional estimators with comparable or better accuracy.
}

\keywords{quantitative MRI, neural networks, Cram\'er-Rao bound, parameter estimation, efficiency}

\maketitle
\footnotetext{\textbf{Word Count: 3217}}

\section{Introduction}\label{sec:intro}
Unbiased parameter estimators are critical for achieving accurate, precise, and reproducible quantitative MRI (qMRI). An unbiased estimator that achieves the Cram\'er-Rao bound (CRB)---the theoretical floor for the variance of an unbiased estimator\cite{Cramer}---is referred to as (statistically) efficient.\cite{Kay1993} While the typical maximum likelihood and least squares estimators used in qMRI---e.g., dictionary matching\cite{Ma2013} and non-linear least squares---are asymptotically efficient with respect to the number of measurements (assuming zero-mean, uncorrelated, homoscedastic Gaussian noise in the case of least squares),\cite{Kay1993,Newey1994,Wu1981} they are \emph{computationally} inefficient, especially when fitting high-dimensional models. Neural networks (NNs) offer significantly reduced fitting time and robustness
\vspace{0mm}at inference, eliminating an important barrier to the clinical adoption of qMRI methods.\cite{Liu2020,Lee2020,Cohen2018a}

Regression neural networks are typically minimum mean squared error (MSE) estimators, i.e., they are trained to minimize the quadratic loss over an empirical dataset. Like other Bayesian approaches,\cite{McGivney2018,Bouhrara2016} their performance is sensitive to the prior (i.e., the training data distribution), where they can achieve a smaller MSE than an efficient estimator by trading off bias for variance. However, this is true only ``on average''---i.e., for parameters close to the mean of the training distribution---and implies that a minimum-MSE NN performs well only if the in vivo data distribution is known \emph{a priori}. This may lead to overly optimistic results when validating the NN's performance in the laboratory setting, with degraded performance in a clinical setting where the qMRI parameters may change in unpredictable ways. While a certain amount of bias might be tolerable for clinical diagnostic use, this is unlikely to be true if the bias varies throughout parameter space, as is typically the case with minimum-MSE NNs. Further, bias impedes inter-method comparability. To minimize this sensitivity to the prior, in this work, we generalize the MSE loss to better control the bias in the NN's estimates.

Though we desire statistical efficiency in the NN, efficient estimators may, in general, not exist or be impractical to find for any given qMRI application, which usually involves a nonlinear multiparametric estimation problem and constraints on parameter space.\cite{Kay1993,Somekh-Baruch2018} The specific biophysical model and measurement scheme employed, as well as areas of parameter space that are inherently difficult-to-estimate, all contribute to this challenge.\cite{Asslander2020} In this work, we therefore soften the requirement that the NN be efficient and instead seek only to promote the associated properties during training; i.e., that the NN have \emph{minimal} bias and a variance \emph{close} to or below the CRB.

\section{Theory}\label{sec:theory}
\subsection{Generalization of the MSE}\label{subsec:gmse}
We begin by writing the multi-variate MSE loss typically used in qMRI, e.g., in DRONE,\cite{Cohen2018a} given by
\begin{equation}
    \mathcal{L}_\mathrm{MSE} \stackrel{\mathrm{def}}{=} \frac{1}{N_sN_p} \sum_{i=1}^{N_s} \big( \hat{\mathbf{x}}(\mathbf{y}_{i}) - \mathbf{x}_i \big)^{T} \big( \hat{\mathbf{x}}(\mathbf{y}_{i})- \mathbf{x}_i \big),
\label{eq:mse}
\end{equation}
where $\mathbf{y} \in \mathbb{C}^{M}$ is the measurement vector, $\mathbf{x}\in \mathbb{R}^{N_p}$ is the ground-truth parameter vector, $\hat{\mathbf{x}} \in \mathbb{R}^{N_p}$ is the NN's estimate, $N_p$ is the number of parameters to estimate, and $N_s$ is the number of samples in the training dataset $\{\mathbf{x}_i,\mathbf{y}_i\}^{N_s}_{i=1}$.

An important limitation of Eq.~\eqref{eq:mse} is that parameters with different units cannot directly be summed together. This can be addressed using a \textit{weighted} loss:
\begin{equation}
    \mathcal{L}_\mathrm{WMSE} \stackrel{\mathrm{def}}{=} \frac{1}{N_sN_p} \sum_{i=1}^{N_s} \big( \hat{\mathbf{x}}(\mathbf{y}_{i}) - \mathbf{x}_i \big)^{T} \mathbf{W}_i \big( \hat{\mathbf{x}}(\mathbf{y}_{i})- \mathbf{x}_i \big),
    \label{eq:wmse}
\end{equation}
where $\mathbf{W}_i \in \mathbb{R}^{N_p \times N_p}$ is a positive semi-definite weighting matrix that tunes each parameter's contribution to the overall loss. 
A diagonal $\mathbf{W}_i$ matrix weights each parameter individually, where weights with the inverse units of their respective parameters render the cost function dimensionless.
The choice $\mathbf{W}_i=\mathbf{I}$ (the identity) reduces Eq.~\eqref{eq:wmse} back to Eq.~\eqref{eq:mse}.

Next, we average over $N_r$ different noise realizations of the measurements (we assume additive white complex Gaussian noise in this work), yielding the weighted MSE loss
\begin{equation}
    \bar{\mathcal{L}}_\mathrm{WMSE} \stackrel{\mathrm{def}}{=} \frac{1}{N_sN_pN_r} \sum_{i=1}^{N_s} \sum_{j=1}^{N_r} \big( \hat{\mathbf{x}}(\mathbf{y}_{ij}) - \mathbf{x}_i \big)^{T} \mathbf{W}_i \big( \hat{\mathbf{x}}(\mathbf{y}_{ij})- \mathbf{x}_i \big).
\label{eq:msew}
\end{equation}
\vspace{0mm}As shown in Appendix~\ref{app1}, this enables decomposition of the loss into the bias and \vspace{0mm}variance as
\begin{equation}
   \bar{\mathcal{L}}_\mathrm{WMSE} = \frac{1}{N_sN_p} \sum_{i=1}^{N_s} \mathrm{Tr}\big\{\mathbf{W}_i \mathrm{Cov}(\hat{\mathbf{x}}_i)\big\} + \big\|\mathrm{Bias}(\hat{\mathbf{x}}_i)\big\|^2_{\mathbf{W}_i},
\label{eq:wmsedecomp}
\end{equation}
\vspace{0mm}where $\mathrm{Cov}(\cdot)$ is the (uncorrected) sample covariance, $\mathrm{Tr}\{\cdot\}$
denotes the matrix trace, and the bias is given by
\begin{gather}
    \mathrm{Bias}(\hat{\mathbf{x}}_i) \stackrel{\mathrm{def}}{=} \hat{\mathbf{\mu}}_i - \mathbf{x}_i \label{eq:bias} \\
    \hat{\mathbf{\mu}}_i \stackrel{\mathrm{def}}{=} \frac{1}{N_r}\sum_{j=1}^{N_r}\hat{\mathbf{x}}(\mathbf{y}_{ij}). \label{eq:sampmean}
\end{gather}
Here, the bias is approximated using the sample mean, so its error w.r.t. the true bias decreases as $\mathcal{O}(1/N_r)$.\cite{Ahn2003}

As suggested by Diskin et. al.,\cite{Diskin2023} we can promote properties similar to an efficient estimator in the trained NN by penalizing the squared bias in addition to the MSE. This leads to the weighted ``bias-constrained error''\cite{Diskin2023} loss
\begin{equation}
    \begin{gathered}
    \bar{\mathcal{L}}_\mathrm{WBCE} \stackrel{\mathrm{def}}{=} \bar{\mathcal{L}}_\mathrm{WMSE} + \frac{\tilde\lambda}{N_sN_p} \sum_{i=1}^{N_s} \big\|\mathrm{Bias}(\hat{\mathbf{x}}_i)\big\|^2_{\mathbf{W}_i} \\
    = \frac{1}{N_sN_p} \sum_{i=1}^{N_s} \mathrm{Tr}\big\{\mathbf{W}_i \mathrm{Cov}(\hat{\mathbf{x}}_i)\big\} + (\tilde\lambda+1)\big\|\mathrm{Bias}(\hat{\mathbf{x}}_i)\big\|^2_{\mathbf{W}_i} ,
    \end{gathered}
    \label{eq:wbce}
\end{equation}
where the non-negative tuning parameter $\tilde\lambda \geq 0$ controls the bias' contribution to the overall cost. While a larger $\tilde\lambda$ generally decreases the bias at the cost of increased variance, it can be difficult to identify a $\tilde\lambda$ in Eq.~\eqref{eq:wbce} that optimally reduces the bias without increasing the variance above the CRB across all estimated parameters in $\hat{\mathbf{x}}$.

\subsection{Variance-Constrained Bias Loss}\label{subsec:varcon}
To achieve this goal, our main contribution is to modify Eq.~\eqref{eq:wbce} to explicitly penalize deviations from an efficient estimator. Here, we primarily consider the CRB-weighting we proposed in Ref.~\citen{Zhang2022}: $\mathbf{W}_i=\mathrm{diagm}(\mathbf{b}_i)^{-1}$,
a matrix with the inverse of the individual parameter's CRBs $\mathbf{b}_i \in \mathbb{R}^{N_p}$ for the measurement $\mathbf{y}_i$ along the main diagonal and 0 elsewhere.
This choice of weighting leads us to formulate the loss
\begin{equation}
    \begin{split}
    \bar{\mathcal{L}}_\mathrm{WVCB}~&\stackrel{\mathrm{def}}{=} \frac{1}{N_sN_p} \sum_{i=1}^{N_s} \sum_{k=1}^{N_p} \bigg( \frac{1}{\mathbf{b}_{ik}} \big( \hat{\mathbf{\mu}}_{ik} - \mathbf{x}_{ik} \big)^2 + \\
    & \lambda \cdot \mathrm{max} \Big(0, \frac{1}{N_r\mathbf{b}_{ik}} \sum_{j=1}^{N_r} \big( \hat{\mathbf{x}}_{k}(\mathbf{y}_{ij})- \hat{\mathbf{\mu}}_{ik} \big)^2 -\delta\Big) \bigg),
    \end{split}
    \label{eq:wvce}
\end{equation}
where the first term penalizes the bias and the second term, where $\lambda \geq 0$, is a variance penalty parameterized by $\delta$; e.g., $\delta=1$ penalizes variances exceeding the CRB.
Note Eq.~\eqref{eq:wvce} reduces to Eq.~\eqref{eq:mse} for $\mathbf{b}_i=\mathbf{1}$ and $N_r=1$, and is equivalent to Eq.~\eqref{eq:wbce} for $\delta=0$, $\lambda=1/(\tilde\lambda+1)$, and $\mathbf{W}_i=\mathrm{diagm}(\mathbf{b}_i)^{-1}$.

Since $\lambda \rightarrow \infty$ is equivalent to imposing a variance constraint, in analogy to the ``bias-constrained error'' term coined in Ref.~\citen{Diskin2023}, we refer to Eq.~\eqref{eq:wvce} as the weighted ``variance-constrained bias loss.'' However, a hard constraint requires the NN to uniformly achieve the CRB across the training set, which may not be possible without increasing the bias. We thus relax this constraint in practice by using a finite $\lambda$.

In the following, we study the use of the proposed NN training strategy for two qMRI applications outlined below.

\section{Methods}\label{sec:methods}
\subsection{Pulse Sequences}
\label{subsec:sequences}
For our first application we used the hybrid-state\cite{Asslander-hsCommPhysics} sequence described in Ref. \citen{Asslander2023} to extract six biophysical parameters ($N_p=6$) of a 2-pool quantitative magnetization transfer (qMT) model,\cite{Henkelman1993,Helms2009,Asslander2021} i.e., the normalized fractional semi-solid spin-pool size $m_0^\mathrm{s}$, the relaxation rates $R_1^\mathrm{f},R_2^\mathrm{f}$ of the free spin-pool, the exchange rate $R_\mathrm{x}$, and the relaxation rates/times $R_1^\mathrm{s},T_2^\mathrm{s}$ of the semi-solid spin-pool. To compute the CRB, we additionally considered three nuisance parameters: the complex-valued scaling $M_0$ and the field inhomogeneities $B_0$ and $B_1^+$, but ignored them for parameter estimation.
This hybrid-state sequence was optimized for a minimal CRB of $m_0^\mathrm{s},R_1^\mathrm{f},R_2^\mathrm{f},R_\mathrm{x},R_1^\mathrm{s}$, and $T_2^\mathrm{s}$ in individual 4s long cycles with antiperiodic boundaries.\cite{Asslander2023,Asslander-hsCommPhysics}

In our second application, we considered the 2D single-slice inversion-recovery MR-fingerprinting FISP (MRF-FISP) sequence\cite{Jiang2015a} designed to estimate a single compartment $T_1$ and $T_2$ ($N_p=2$). We additionally considered $M_0$ only to compute the CRB. We used $14.6\pi$/voxel spoiling along the slice select direction, 1ms sinc-pulses with a time-bandwidth product of 4, TR=10ms, TE=5ms, and TI=20ms (time between adiabatic inversion pulse and first sinc-pulse).

\subsection{Training Data}
\label{subsec:trainingdata}
For the qMT model, we simulated a dictionary of approximately 600,000 fingerprints with the generalized Bloch framework\cite{Asslander2021} using randomly generated parameters drawn from a mixture of Gaussian distributions centered around the values expected to be measured in vivo. Gaussians were preferred to uniform distributions to capture the high-dimensional parameter space with comparably few samples while still approximating the distribution expected in vivo.
We heuristically chose 80\% of fingerprints to have parameters typical for gray and white matter ($m_0^\mathrm{s} \in \mathcal{N}(0.2, 0.2)_0$ ($\mathcal{N}(m,d)$ denoting a normal distribution with mean $m$ and standard deviation $d$ and scripts denoting truncation limits), $T_1^f \in \mathcal{N}(3, 2)$s, $R_2^\mathrm{f} \in \mathcal{N}(15, 10)$s$^{-1}$, $R_\mathrm{x} \in \mathcal{N}(30, 10)$s$^{-1}$, $R_1^\mathrm{s} \in \mathcal{N}(4, 2)$s$^{-1}$, $T_2^\mathrm{s} \in \mathcal{N}(10, 3)_5^{20}\upmu$s$^{-1}$), 10\% parameters typical for fat ($m_0^\mathrm{s} \in \mathcal{N}(0.1, 0.1)_0$, $T_1^f \in \mathcal{N}(400, 75)$ms, $T_2^f \in \mathcal{N}(100, 20)$ms, $R_\mathrm{x} \in \mathcal{N}(30, 10)$s$^{-1}$, $R_1^\mathrm{s} \in \mathcal{N}(4, 2)$s$^{-1}$, $T_2^\mathrm{s} \in \mathcal{N}(10, 3)_5^{20}\upmu$s$^{-1}$), and 10\% parameters that are typical for CSF ($m_0^\mathrm{s}=0$, $T_1^f \in \mathcal{N}(4, 0.5)$s, $T_2^f \in \mathcal{N}(2, 0.25)$s).\cite{Asslander2023,Bojorquez2017,Stanisz2005} Field inhomogeneities were simulated with $B_0 \in \mathcal{U}_{[-\pi/\mathrm{TR},\pi/\mathrm{TR}]}$ (uniform distribution) and $B_1^+ \in \mathcal{N}(0.9, 0.3)_{0.6}^{1.2}$. We used an SVD of the full simulated dictionary to compute a temporal subspace of rank 15 and compress the dictionary to emulate the typical subspace reconstruction measurement process;\cite{McGivney2014,Tamir2017,Asslander2018,Zhao2018,Mao2023} i.e., $\mathbf{y} \in \mathbb{C}^{15}$ in this case. For practical reasons, we used only $N_s=1.84\cdot10^5$ samples for training the NN, of which 20\% were reserved for testing.

For the MRF-FISP sequence, we computed a dictionary of $N_s=281,250$ fingerprints with Bloch simulations spanning grey/white matter, fat, and CSF values at 3T: $T_1$ in the ranges [500:2:1500]ms, [250:2.4:550]ms, [3000:16:5000]ms (min:stepsize:max) and $T_2$ in the ranges [10:0.38:200]ms, [60:0.65:140]ms, [1500:8.1:2500]ms, respectively.\cite{Jiang2015a,Bojorquez2017} The transmit field strength was assumed to be uniform, i.e., $B_1^+ =1$. We accounted for the slice profile by taking the complex average of 1324 isochromats with 600 isochromats uniformly distributed with $14.6\pi$ phase between the FWHMs of the small flip angle approximated slice profile.\cite{Malik2016} For CSF's long relaxation times, we instead simulated a total of 5300 isochromats across the slice profile. This dictionary was used for computing the rank 10 temporal subspace, dictionary compression, and network training.

\subsection{Loss Functions} \label{subsec:losses}
We consider two variants of the loss in Eq.~\eqref{eq:wvce}:
\begin{enumerate}[leftmargin=10mm]
    \item MSE-CRB:\cite{Zhang2022}~~$\lambda=1$,~~$\delta=0$,~~$N_r=1$
    \item Bias-Reduced:~~$\lambda>0$,~~$\delta=1$,~~$N_r=200$.
\end{enumerate}
Strategy (1) is a state-of-the-art approach that improves upon the MSE (Eq.~\eqref{eq:mse}) by accounting for variations in scale across the different parameters and ensuring robustness to difficult-to-estimate parameters during training. 
While Eq.~\eqref{eq:wvce} shows that the loss would be 0 for an efficient NN, the CRB-weighting does not in itself ensure efficiency. Hence, strategy (2) improves on (1) by introducing averaging over multiple noise realizations to enable finer control over the NN's bias and variance properties. We empirically determined the suitable regularization strength to be $\lambda=1$ for both qMRI applications. While $\delta=1$ is chosen for simplicity, $\delta=1.01$ could also be used to account for the variance of the sample variance for $N_r=200$.\cite{Ahn2003}

\begin{figure*}[tbp]
    \includegraphics[width=0.975\textwidth]{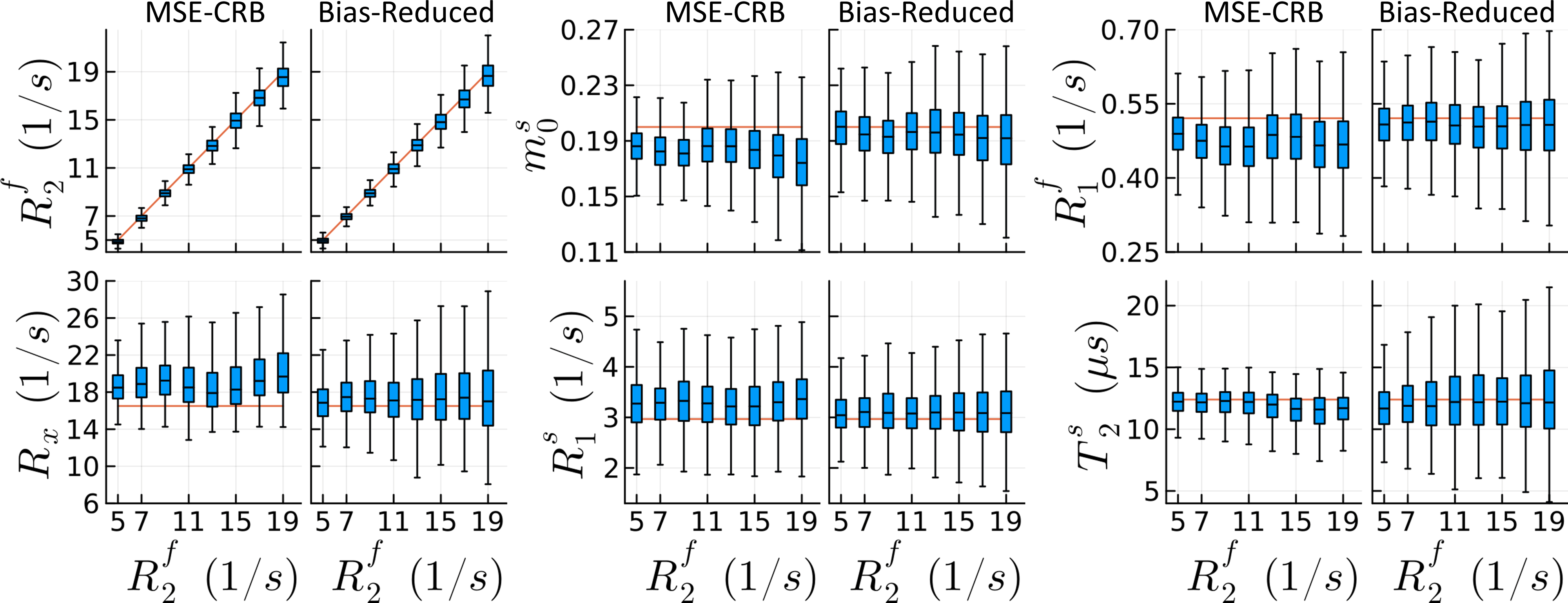}
    \caption{Boxplot comparison of simulated qMT parameter fits with networks trained using a state-of-the-art (MSE-CRB or Cram\'er-Rao bound weighted mean squared error criterion\cite{Zhang2022}) and the proposed Bias-Reduced loss, assuming SNR $=20$. As an example, we vary only $R_2^\mathrm{f}$ (the free spin pool's transverse relaxation rate) while keeping all other parameters constant (red reference lines are the ground-truth). The proposed strategy significantly reduces the variable bias in all other parameters throughout parameter space except $T_2^\mathrm{s}$ ($p<10^{-3}$ using Welch's unequal variances $t$-test).
    }
    \label{fig:mt_bias_1d}
\end{figure*}

\subsection{Network Architecture and Training}
For the qMT model, we use a slightly modified version of the NN architecture we used in Ref.~\citen{Zhang2022}: 11 fully connected layers with skip connections and batch normalization, a maximum layer width of 1024, and an input layer where $\mathbf{y}$ (the 15 temporal coefficients) is split into real and imaginary parts, with a total of 2,187,138 trainable parameters. The outputs are also constrained using ReLUs capped at the maximum values expected in vivo (note they could also be clamped to the minimum). We train the NNs using the Rectified ADAM optimizer,\cite{Liu2019} a learning rate of $10^{-4}$, and a batch size of 256.

For MRF-FISP, we adapt the original DRONE architecture,\cite{Cohen2018a} retaining the same fully connected 3 layers but modifying the input layer to be the real and imaginary parts of $\mathbf{y}$ (the 10 temporal coefficients), using ReLU activations, incorporating batch normalization, and constraining the output values using ReLUs capped at the maximum values expected in vivo (5 for both $T_1$ and $T_2$ respectively), with a total of 98,402 trainable parameters. We train the NNs using the ADAM optimizer,\cite{Kingma2014} a learning rate of $10^{-4}$, and a batch size of 256.

For each pulse sequence, we first trained a NN with $\lambda=0$ and $N_r=1$ to convergence. This network is used to initialize the Bias-Reduced NNs trained for 500 epochs with $N_r=200$, $\lambda=1$ and $\delta=1$. From the same initialization, we trained an NN with the MSE-CRB loss for a further 500 epochs with a batch size of 51200 for fair comparison to the state-of-the-art.

To ensure robustness to variable noise levels, we added white complex Gaussian noise $\epsilon \sim \mathcal{N}(0,\sigma)$ to the training data for $\mathrm{SNR} \stackrel{\mathrm{def}}{=} |M_0|/\sigma \in (10,50)$. A random SNR is selected for each measurement $\mathbf{y}_i$ at each training epoch, which is used to generate the $N_r=200$ noise realizations.

\subsection{Simulation Experiments} \label{subsec:sim}
To study the bias and variance properties of the proposed NNs, we simulate measurements for both our applications as described in Section~\ref{subsec:trainingdata} but instead corrupt with white complex Gaussian noise $N_r=1000$ times. For the qMT sequence, we compared our NNs to non-linear least squares (NLLS) using the Levenberg-Marquardt algorithm,\cite{Marquardt1963} initialized with the ground-truth. For MRF-FISP, we compared to dictionary matching, a discretized maximum likelihood estimator commonly used in the MRF literature.\cite{Ma2013}


\subsection{In Vivo Experiments}
To evaluate the effect of the proposed NN estimators in vivo, we scanned two healthy volunteers on a 3T Prisma system (Siemens, Erlangen, Germany) with a 32-channel head coil after obtaining informed consent in agreement with our IRB's requirements.
For the qMT sequence, we scanned the whole brain of one subject using 3D radial koosh-ball k-space sampling with a 2D golden means pattern\cite{Chan2009} reshuffled to improve k-space coverage and minimize eddy current artifacts\cite{Flassbeck2023} with a 256mm isotropic FOV and 1.24mm isotropic effective resolution, repeating the hybrid-state sequence for 12\,min scan time.
For MRF-FISP, we acquired a single axial slice in subject two's brain using 2D golden-angle radial k-space sampling instead of spirals, with an FOV=$256\times256$mm, voxel size $1\times1\times4$mm, and 10 cycles of the sequence (10 radial spokes per frame), for 3.4\,min of scan time.

For both sequences, we use the low-rank inversion subspace reconstruction approach\cite{McGivney2014,Asslander2018} to reconstruct coefficient images directly in the subspace.
A locally low-rank penalty \cite{Trzasko2011,Zhang2015} is used to suppress artifacts for the qMT sequence (which we note modifies the noise distribution from the assumed white complex Gaussian noise during NN training).
The reconstructed coefficients are used voxel-by-voxel as NN inputs to estimate the biophysical parameters.

\begin{figure*}[tbp]
    \centering
    \includegraphics[width=0.895\textwidth]{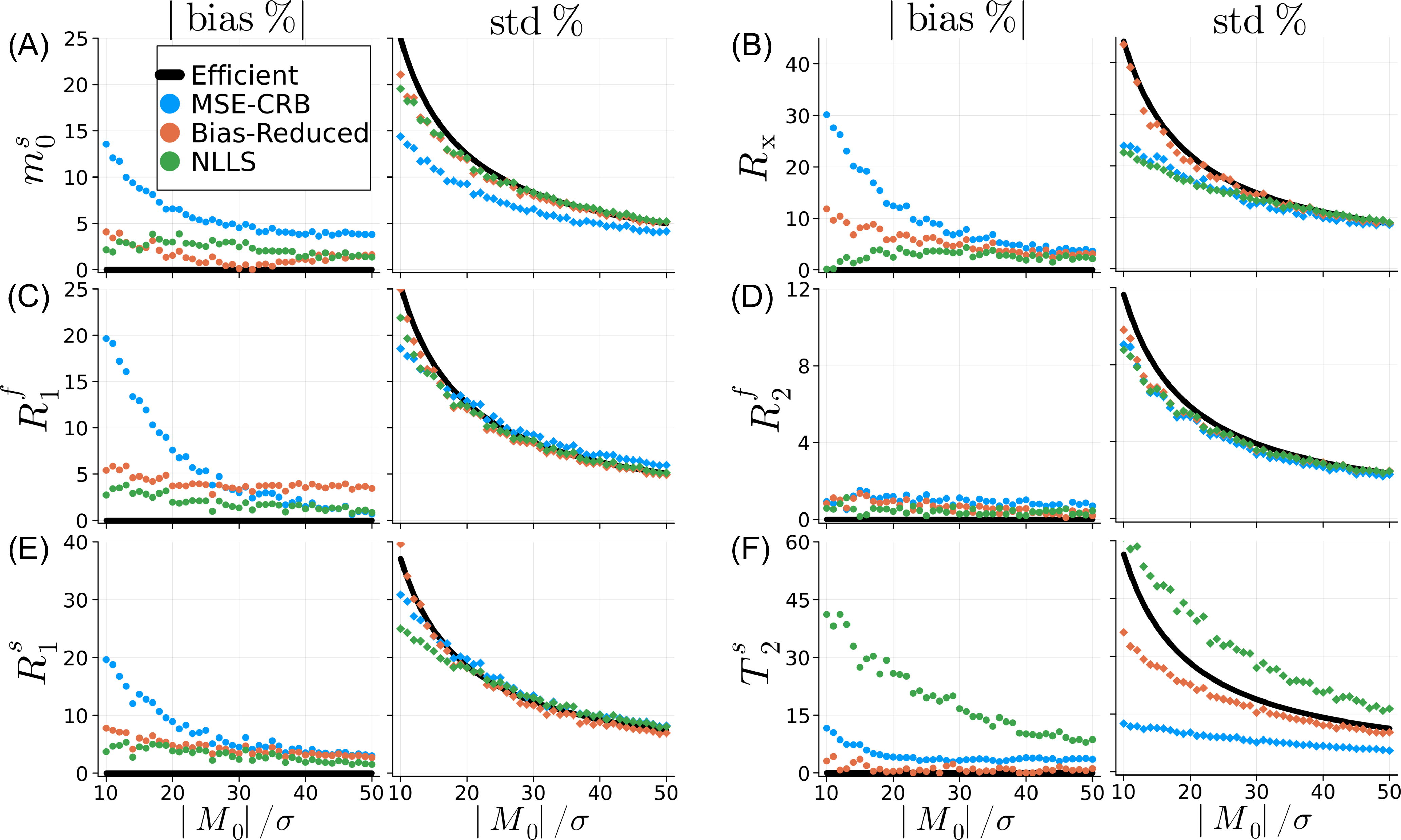}
    \caption{Normalized bias and standard deviation of the magnetization transfer parameter estimates as a function of SNR ($|M_0|/\sigma$). The estimates are based on simulations using typical white matter values ($m_0^\mathrm{s}=0.2$, $R_1^\mathrm{f}=0.52/s$, $R_2^\mathrm{f}=12.9/s$, $R_\mathrm{x}=16.5/s$, $R_1^\mathrm{s}=2.97/s$, $T_2^\mathrm{s}=12.4\mu s$),\cite{Asslander2023} which are used for normalization. We compare neural networks trained using the MSE-CRB\cite{Zhang2022} and proposed Bias-Reduced ($\lambda=1$) losses to non-linear least squares (NLLS) and a hypothetical efficient estimator, which has zero bias and variance equal to the Cram\'er-Rao bound. The proposed strategy is similar in performance to NLLS in all parameters except $T_2^\mathrm{s}$, where it more closely matches an efficient estimator. This analysis, repeated for grey matter values, is shown in Sup.~Fig.~\ref{sfig:mt_bv}.}
    \label{fig:mt_bv}
\end{figure*}

\section{Results}\label{sec:results}
Fig.~\ref{fig:mt_bias_1d} visualizes how the proposed strategy reduces the NN's variable bias along one axis in parameter space. While the MSE-CRB NN can correctly estimate the simulated changes in $R_2^\mathrm{f}$, most other parameters exhibit bias that is also variable throughout parameter space. The Bias-Reduced strategy yields overall reduced bias, at the cost of overall increased variance (particularly for $T_2^\mathrm{s}$). However, we observe for the Bias-Reduced NN that the only major effect of increasing $R_2^\mathrm{f}$ is an increased variance of all parameters' estimates---consistent with the expected increase in CRB.

\begin{figure*}[tbp]
    \centering
    \includegraphics[width=0.925\textwidth]{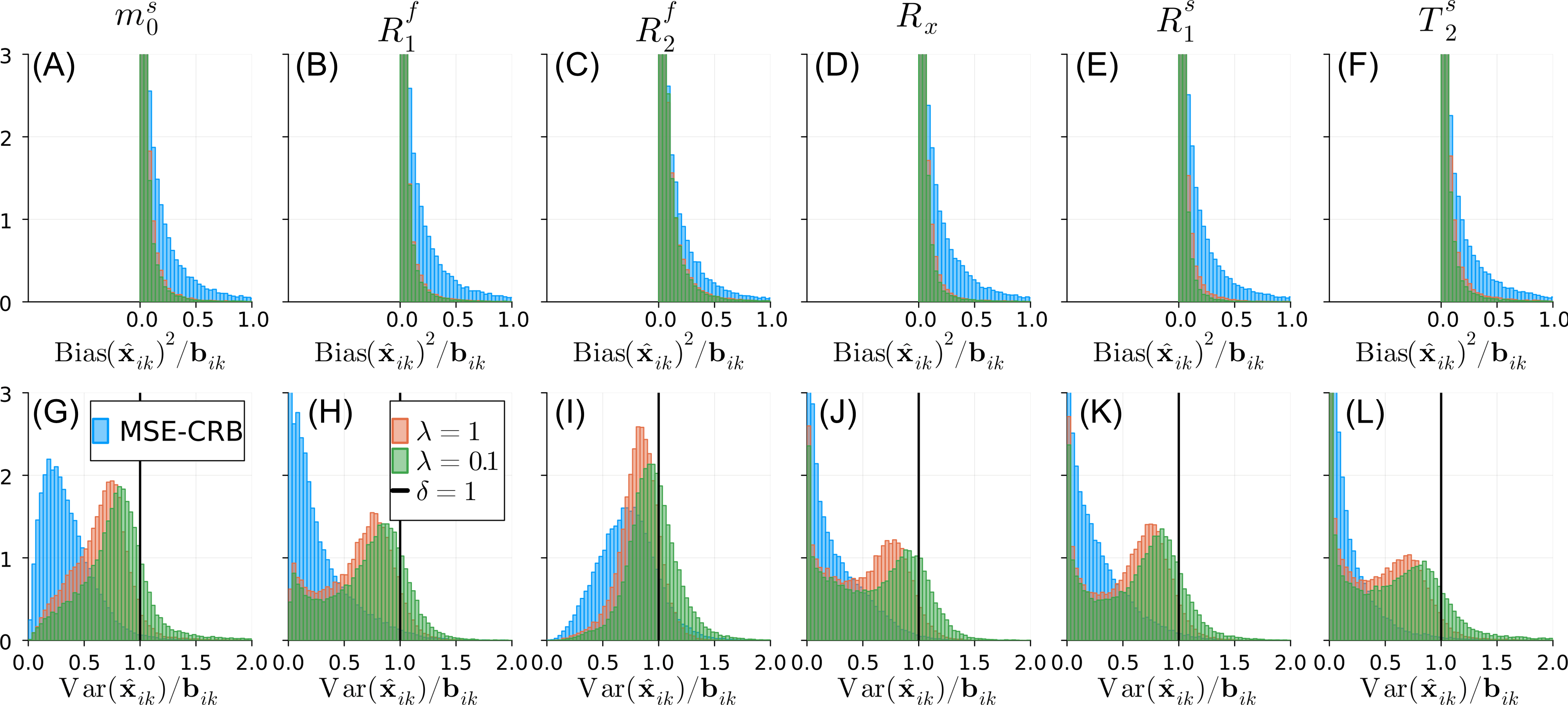}
    \caption{Normalized histograms of the CRB-weighted ($\mathbf{b}_{ik}$) squared bias (A--F) and variance (G--L) of the magnetization transfer parameter estimates $\{\hat{\mathbf{x}}_{ik}\}$, where $i$ indexes across the test set (where each fingerprint has a random SNR) and $k$ indexes across parameters. Note the scaling of the y-axis truncates the left-most bins in each subplot. The MSE-CRB\cite{Zhang2022} loss (blue) offers the lowest variance but the highest bias overall in comparison to the proposed Bias-Reduced strategies $\lambda=1$ (red) and $\lambda=0.1$ (green). A smaller $\lambda$ ($\lambda=0.1$) reduces the overall bias slightly at the outsized cost of an increased proportion of fingerprints exceeding the CRB ($\delta=1$; Eq.~\eqref{eq:wvce}). A comparison to Eq.~\eqref{eq:wbce} is shown in Sup.~Fig.~\ref{sfig:mt_hist}.}
    \vspace{-5mm}
    \label{fig:mt_hist}
\end{figure*}

\begin{figure*}
    \centering
    \includegraphics[width=0.925\textwidth]{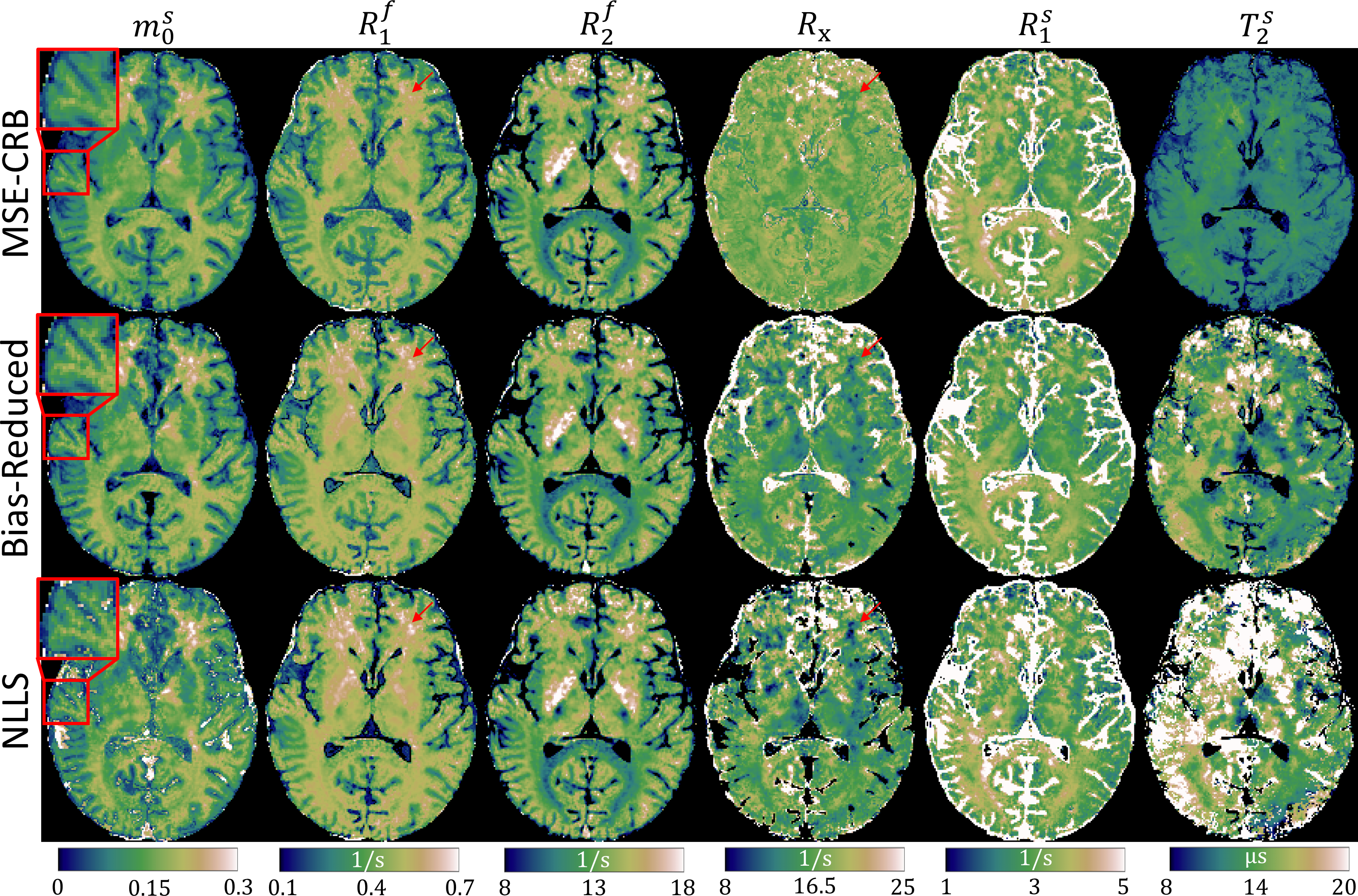}
    \caption{In vivo magnetization transfer parameter maps fitted with the MSE-CRB\cite{Zhang2022} and proposed Bias-Reduced neural networks in comparison to a non-linear least squares (NLLS) reference. The Bias-Reduced network offers the highest visual contrast in $m_0^\mathrm{s}$ (magnifications) and has improved consistency with NLLS in all parameters (particularly $R_1^\mathrm{f}$ and $R_\mathrm{x}$, red arrows) except for $T_2^\mathrm{s}$, consistent with Fig.~\ref{fig:mt_bv}F.}
    \vspace{-2mm}
    \label{fig:mt_iv}
\end{figure*}

\begin{figure*}[tbp]
    \centering
    \includegraphics[width=0.95\textwidth]{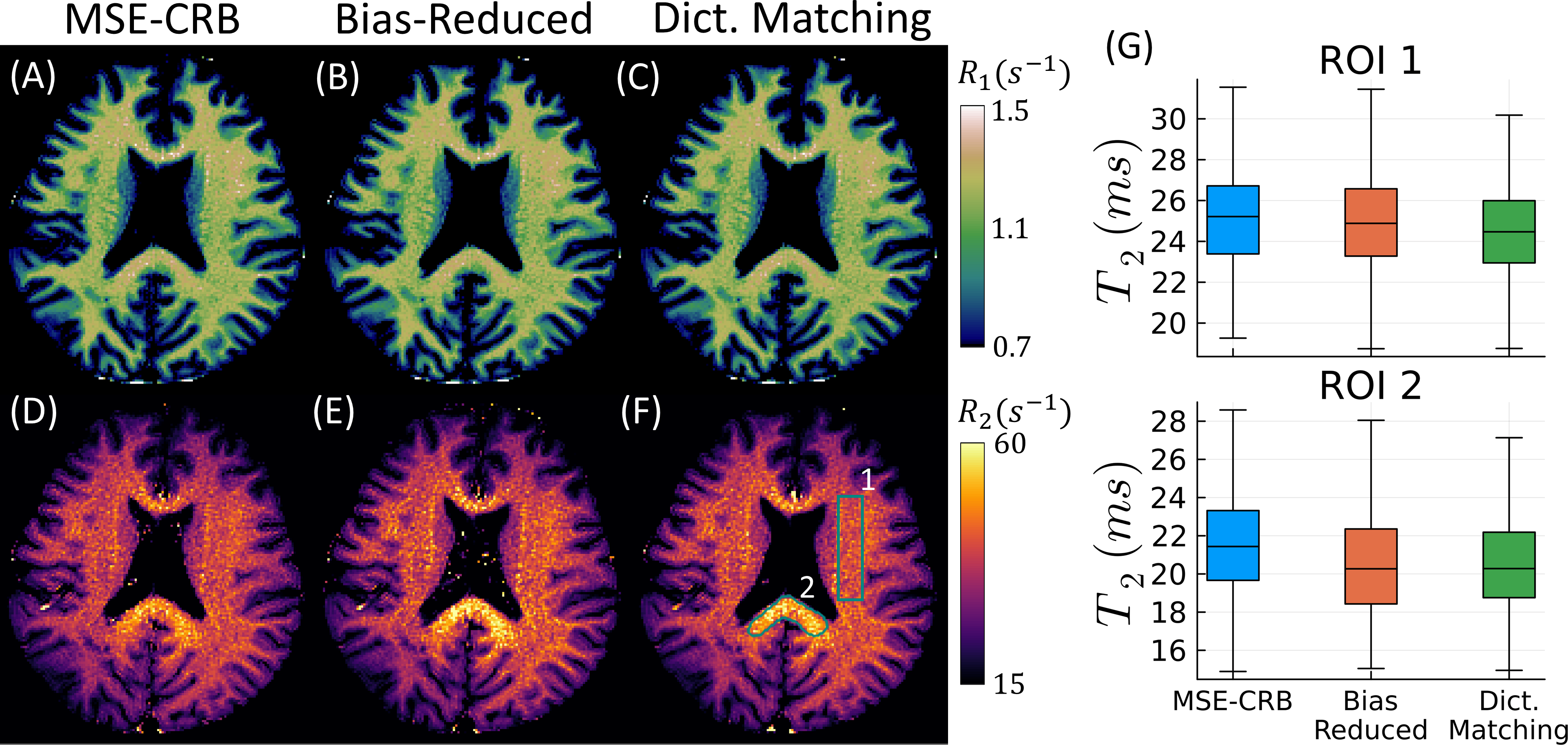}
    \caption{In vivo $1/T_1$ (A--C) and $1/T_2$ (D--F) maps acquired using the MRF-FISP sequence and fitted using NNs trained with two different strategies in comparison to a dictionary-matching-based reference. (G) analyzes the $T_2$ values within the two white matter ROIs drawn in (F), where outliers are not plotted. The Bias-Reduced NN yields more similar parameter maps to those of dictionary matching, but with the benefit of improved computational efficiency. Similar accuracy and precision to dictionary matching is also observed in simulation (Sup.~Figs.~\ref{sfig:fisp_bv_t2}--\ref{sfig:fisp_bv_t1}).}
    \label{fig:fisp_iv}
\end{figure*}

Fig.~\ref{fig:mt_bv} analyzes the simulated bias and variance as a function of SNR for a single point in parameter space corresponding to white matter.\cite{Asslander2023} In several parameters, e.g., $m_0^\mathrm{s}$ and $T_2^\mathrm{s}$, the MSE-CRB NN achieves the lowest variance for all SNR values at the cost of increased bias. The proposed loss reduces the bias for most parameters and SNR levels (except, e.g., $R_1^\mathrm{f}$ at high SNR) and the resulting variance, while larger, more closely follows the CRB. Notably, the Bias-Reduced NN outperforms NLLS in $T_2^\mathrm{s}$ in both bias and variance, despite the latter's initialization with the ground-truth (initialization is inapplicable for NNs at inference).

Fig.~\ref{fig:mt_hist} evaluates the impact of $\lambda$ for the Bias-Reduced loss across the entire test set where each fingerprint has a random SNR. The proposed strategy with $\lambda=1$ significantly reduces the overall bias for all parameters relative to the MSE-CRB criterion ($p<10^{-6}$ using the Wilcoxon signed-rank test)---notably, including $T_2^\mathrm{s}$---with a variance less than or equal to the CRB for the majority of fingerprints. Smaller $\lambda$ values lead to further reduced overall bias at the cost of increased overall variance. Similar analyses of the NN's performance for parameter and SNR values beyond the ranges used for training are shown in Sup.~Figs.~\ref{sfig:mt_robust} and \ref{sfig:mt_hist_highsnr}, respectively. Sup.~Fig.~\ref{sfig:mt_hist} shows that, while a NN trained using Eq.~\eqref{eq:wbce} has a similar bias, the proposed strategy has more uniform variance properties across all estimated parameters.

Fig.~\ref{fig:mt_iv} investigates the effect the proposed strategy has on in vivo parameter fits. In comparison to the MSE-CRB, the Bias-Reduced strategy yields improved visual contrast in $m_0^\mathrm{s}$, which is likely a result of a reduced bias towards the white matter prior in the training data. There is greater correspondence with the reference NLLS maps in the harder-to-estimate parameters---particularly $R_\mathrm{x}$---though substantial differences remain in $T_2^\mathrm{s}$, which is line with Fig.~\ref{fig:mt_bv}F.

Similar results are seen in simulation and in vivo using the FISP sequence, though less pronounced. Sup.~Fig.~\ref{sfig:fisp_bv_t2} shows that the proposed Bias-Reduced strategy has the lowest overall $T_2$ bias throughout parameter space in comparison to the MSE-CRB NN and dictionary matching. As seen in Fig.~\ref{fig:fisp_iv}, all three estimators perform similarly in vivo with respect to $T_1$, but the Bias-Reduced strategy produces the $T_2$ maps most similar to dictionary matching. This is quantified in the boxplots in Fig.~\ref{fig:fisp_iv}G, which, in line with Fig.~\ref{fig:mt_bias_1d}, demonstrates that the proposed Bias-Reduced strategy reduces ROI-dependent bias in vivo. Specifically, bias is reduced for the extreme $T_2$ values measured in the splenium, which have a comparably large CRB (cf. Fig.~\ref{sfig:fisp_bv_t2}G). In Sup.~Fig.~\ref{sfig:fisp_iv}, we show that the improvement offered by the proposed strategy is not simply due to averaging over multiple noise realizations.

\section{Discussion} \label{sec:disc}
We propose a simple training loss enabling control over the bias and variance properties of NN parameter estimators. We show empirically in two qMRI applications that the proposed loss reduces the bias in comparison to the traditional MSE loss while keeping the variance close to the CRB. Such NNs are beneficial for developing and validating new qMRI biomarkers, particularly for advanced biophysical models that attempt to move beyond the standard Bloch equations; e.g., myelin water imaging, magnetization transfer, and diffusion. The proposed NNs are also expected to be more robust to pathology, where deviations from the prior are unpredictable and cannot be known \textit{a priori}.

In this article, we considered the CRB-weighted MSE loss, which emphasizes different areas of parameter space depending on their ease of estimation. This, in addition to the training data distribution, can be thought of as priors that affect the NN's generalization capabilities. The MSE-CRB-trained NN learns to minimize the loss by reducing the variance at the cost of bias towards the prior. The proposed approach reduces the impact of these priors while promoting the properties of an efficient estimator throughout parameter space. 
In general, we expect that the proposed method is more beneficial for high-dimensional qMRI models with difficult-to-estimate parameters that cannot feasibly be sampled on a uniform grid to generate training data, e.g., the Standard Model of diffusion.\cite{Novikov2019}

We obtained similar results when employing NN architectures of varying sizes, and postulate that the NN only needs to have sufficient expressivity to capture the complexity of the training data and estimation problem. For example, recent work regarding the ``interpolation point'' of NNs suggests that the number of trainable parameters should be greater than the number of examples times the number of estimated parameters.\cite{Belkin2019} However, a thorough investigation of the best choice of architecture and nonlinear activation functions is outside this article's scope.

As ground-truth parameters are often unknown in vivo, our approach helps reduce uncertainty about the quality of the NN's estimates. As we view the regression NN through the lens of an estimator, our approach is related to other work surrounding quantifying uncertainty in NNs, e.g., by estimating the variance of the NN's predictions in a Bayesian framework.\cite{Lambert2022,Abdar2021,Jalal2021,Luo2023} By focusing on promoting the properties of efficient estimators, our approach avoids the limitations of training NNs to reproduce the fallible estimates of traditional estimators.\cite{Liu2020,Lee2020,Cohen2018a} While self-supervised methods also encourage unbiasedness to some degree,\cite{Luu2023} they are only effective primarily for the parameters with the largest signal derivatives. Encoding bias reduction into the NN's weights during training also reduces the need to apply a computationally expensive bias correction after estimation.\cite{Kosmidis2014,Whitaker2022}

An important limitation lies in our adoption of the common assumption that the signal is perfectly described by the biophysical model plus white complex Gaussian noise. Unmodeled biophysical effects in the experimental data, however, are usually not Gaussian distributed. Imaging artifacts from various sources,\cite{AjaFernndez2016} advanced imaging techniques such as parallel imaging,\cite{Varadarajan2015,Bouhrara2018} and regularized image reconstruction\cite{Fessler2002} can further alter the residuals' distribution. Our approach does not reduce bias resulting from this ``data mismatch,'' which applies to all estimators considered in this article.

There are several interesting avenues for future work. 
Eq.~\eqref{eq:wmsedecomp} weights both the bias and covariance terms with the same weighting matrix $\mathbf{W}$ and the employed CRB-weighting de-emphasizes difficult-to-estimate areas of parameter space during training. If the bias needs to be further reduced in areas where the CRB is large, one could design a $\mathbf{W}$ that only normalizes the different parameters by their average value within the training dataset.
Further, the MSE-CRB strategy considers only the individual parameter's variances and ignores the off-diagonal elements of the covariance matrix in Eq.~\eqref{eq:wmsedecomp}, which is equivalent to considering the estimation of each parameter separately. However, this assumption may not necessarily hold true for the employed network architecture, and future work will consider separate NNs trained to regress each parameter individually. Eq.~\eqref{eq:wmsedecomp} also offers the flexibility to design a non-diagonal $\mathbf{W}$ to explicitly penalize the covariances, which could be beneficial for a joint statistical analysis of the parameter estimates.

\section{Conclusion}\label{sec:conc}
A tunable generalization of the MSE loss enables training NNs that are more similar to efficient estimators than those trained with the traditional MSE loss. The proposed NNs are well-suited for the development and validation of new quantitative biomarkers.

\section*{Acknowledgments}
This work was supported by NIH grants F30~AG077794, R01~NS131948, T32~GM136573, and was performed under the rubric of the Center for Advanced Imaging Innovation and Research (CAI2R), an NIBIB National Center for Biomedical Imaging and Bioengineering (P41~EB017183).

\section*{Data Availability Statement}\label{sec:DataAvailibility}
\textit{Julia} code to train neural networks for the qMT model are uploaded to the Github repository \href{https://www.github.com/andrewwmao/BiasReducedNetworks}{@andrewwmao/BiasReducedNetworks}. Example scripts are also provided to reproduce Figs.~\ref{fig:mt_bias_1d}, \ref{fig:mt_bv}, and \ref{fig:mt_iv}, except for non-linear least squares fitting, the code for which is already available at \href{https://www.github.com/JakobAsslaender/MRIgeneralizedBloch.jl}{@JakobAsslaender/MRIgeneralizedBloch.jl}.

\appendix
\section{Derivation of the Weighted MSE Loss}\label{app1}
For simplicity, here we consider only the sample mean over the noise realizations in Eq.~\eqref{eq:msew}, ignoring the sum over different samples within the training set. Now,
\begin{equation*}
    \begin{aligned}
    &\frac{1}{N_r} \sum_{j=1}^{N_r} ( \hat{\mathbf{x}}_j - \mathbf{x} )^{T} \mathbf{W} ( \hat{\mathbf{x}}_j - \mathbf{x} ) \\
    &= \frac{1}{N_r} \sum_{j=1}^{N_r} \mathrm{Tr} \big\{ \mathbf{W} (\hat{\mathbf{x}}_j - \mathbf{x}) (\hat{\mathbf{x}}_j - \mathbf{x})^T \big\} \\
    &\stackrel{i}{=} \mathrm{Tr} \Big\{ \mathbf{W} \cdot \frac{1}{N_r} \sum_{j=1}^{N_r} (\hat{\mathbf{x}}_j - \mathbf{x}) (\hat{\mathbf{x}}_j - \mathbf{x})^T \Big\} \\
    &\stackrel{ii}{=} \mathrm{Tr} \Big\{ \mathbf{W} \cdot \frac{1}{N_r} \sum_{j=1}^{N_r} \big(\hat{\mathbf{x}}_j - \hat{\mathbf{\mu}} + \mathrm{Bias}(\hat{\mathbf{x}})\big) \big(\hat{\mathbf{x}}_j - \hat{\mathbf{\mu}} + \mathrm{Bias}(\hat{\mathbf{x}})\big)^T \Big\} \\
    &= \mathrm{Tr} \Big\{ \mathbf{W} \cdot \frac{1}{N_r} \sum_{j=1}^{N_r} \big( (\hat{\mathbf{x}}_j - \hat{\mathbf{\mu}}) (\hat{\mathbf{x}}_j - \hat{\mathbf{\mu}})^T + \mathrm{Bias}(\hat{\mathbf{x}}) \cdot \mathrm{Bias}(\hat{\mathbf{x}})^T\\
    &\hspace{23mm} + (\hat{\mathbf{x}}_j - \hat{\mathbf{\mu}}) \cdot \mathrm{Bias}(\hat{\mathbf{x}})^T + \mathrm{Bias}(\hat{\mathbf{x}}) (\hat{\mathbf{x}}_j - \hat{\mathbf{\mu}})^T \big) \Big\} \\
    &\stackrel{iii}{=} \mathrm{Tr} \big\{ \mathbf{W} \mathrm{Cov}(\hat{\mathbf{x}}) \big\} + \mathrm{Tr} \big\{ \mathbf{W} \mathrm{Bias}(\hat{\mathbf{x}}) \cdot \mathrm{Bias}(\hat{\mathbf{x}})^T \big\} \\
    &= \mathrm{Tr} \big\{ \mathbf{W} \mathrm{Cov}(\hat{\mathbf{x}}) \big\} + \mathrm{Bias}(\hat{\mathbf{x}})^T\mathbf{W}\mathrm{Bias}(\hat{\mathbf{x}}) \\
    &\stackrel{iv}{=} \mathrm{Tr} \big\{ \mathbf{W} \mathrm{Cov}(\hat{\mathbf{x}}) \big\} + \big\|\mathrm{Bias}(\hat{\mathbf{x}})\big\|^2_\mathbf{W},
    \end{aligned}
\end{equation*}
where (\emph{i}) uses the linearity of the trace, (\emph{ii}) follows from inserting Eq.~\eqref{eq:bias}, (\emph{iii}) uses the fact that $\frac{1}{N_r} \sum_{j=1}^{N_r} (\hat{\mathbf{x}}_j - \hat{\mathbf{\mu}}) = \mathbf{0}$ and the definition of the (uncorrected) sample covariance as $\mathrm{Cov}(\hat{\mathbf{x}}) \stackrel{\mathrm{def}}{=} \frac{1}{N_r} \sum_{j=1}^{N_r} (\hat{\mathbf{x}}_j - \hat{\mathbf{\mu}}) (\hat{\mathbf{x}}_j - \hat{\mathbf{\mu}})^T$, and (\emph{iv}) uses the notation $\|\mathbf{a}\|^2_\mathbf{W} \stackrel{\mathrm{def}}{=} \mathbf{a}^T \mathbf{W} \mathbf{a}$.

Note that since Eq.~\eqref{eq:msew} is written as a sample mean---which is necessary for training a network in practice---it only approximates the mean squared error written as an expectation over the noise distribution. The derivation shown here holds within this approximation of the expectation using a sample mean.

\bibliography{MRM-AMA}

\section*{Supporting information}
The following supporting information is available as part of the online article:

\vskip\baselineskip
\noindent
\textbf{Figure S1.}
Comparison of the MSE-CRB and Bias-Reduced NN's CRB-weighted squared bias and variance for 500 qMT fingerprints (each with a random SNR) randomly sampled from a mixture of Gaussian distributions truncated at non-physical values (e.g., constraints like $0 \leq m_0^\mathrm{s} \leq 1$ are still imposed). Fingerprints that are outside the cutoff ranges of the training data distribution in any parameter are colored red, and otherwise blue. The x-axis shows the Euclidean distance from the mean of the training distribution, weighted by the standard deviations (e.g., calculated from z-scores of the individual parameters), and thus (approximately) follows the Chi distribution. The dashed black line corresponds to the Cram\'er-Rao Bound. For both networks, the bias is generally higher for fingerprints outside the training distribution. The proposed training strategy reduces the overall bias for fingerprints both in and outside of the distribution.

\noindent
\textbf{Figure S2.}
Repetition of Fig.~\ref{fig:mt_bv} showing the normalized absolute percent bias and percent standard deviation of the 2-pool qMT parameters in grey matter ($m_0^\mathrm{s}=0.091,R_1^\mathrm{f}=0.37/s,R_2^\mathrm{f}=11.9/s,R_\mathrm{x}=20.5/s,R_1^\mathrm{s}=3.17/s,T_2^\mathrm{s}=11.7\mu s$)\cite{Asslander2023} as a function of SNR ($|M_0|/\sigma$) for neural networks trained using the MSE-CRB\cite{Zhang2022} and proposed Bias-Reduced losses in comparison to non-linear least squares (NLLS) and a hypothetical efficient estimator. Here, similarly to white matter, the proposed strategy performs similarly to NLLS in all parameters except $T_2^\mathrm{s}$, where the performance is more in line with an efficient estimator.

\noindent
\textbf{Figure S3.}
Repetition of Fig.~\ref{fig:mt_hist} where all fingerprints in the test set have a random SNR higher than the range of SNRs seen during training. In this case, the bias is overall higher for both networks. While this suggests somewhat impaired generalization of the employed NN architecture,\cite{Mohan2020} it is also consistent with normalization by smaller Cram\'er-Rao bounds---which account for the decreased noise level---and an expected floor to the accuracy of the NN estimator that is related purely to measurement noise. The proposed strategy for reducing bias still holds outside of the training range of SNRs, albeit somewhat less for $R_2^\mathrm{f}$.

\noindent
\textbf{Figure S4.}
Repetition of Fig.~\ref{fig:mt_hist} comparing NNs trained with the MSE-CRB, the proposed Bias-Reduced and the bias-constrained loss (Eq.~\eqref{eq:wbce}) with an optimized lambda. While the bias-constrained approach has similar bias to the Bias-Reduced strategy, it has less uniform variance properties across all estimated qMT parameters with a longer tail past the $\delta=1$ line.

\noindent
\textbf{Figure S5.}
Normalized bias and standard deviation of the FISP-based $T_2$ estimates as a function of $T_1$ and $T_2$ using NNs trained with two different strategies in comparison to a dictionary-matching-based reference (C,F). (A,D) The Cram\'er-Rao bound weighted mean squared error (MSE-CRB).\cite{Zhang2022} (B,E) The Bias-Reduced strategy achieves the lowest overall bias throughout parameter space with a similar variance to the CRB reference (G). The green circle marks the average white matter values measured in vivo (Fig.~\ref{fig:fisp_iv}).

\noindent
\textbf{Figure S6.}
Repetition of Sup.~Fig.~\ref{sfig:fisp_bv_t2} showing the normalized absolute percent bias and percent standard deviation of the FISP-based $T_1$ estimates instead. In this case, the performance is similar between NNs trained using both strategies and the reference.

\noindent
\textbf{Figure S7.}
Comparison of in vivo FISP $1/T_1$ and $1/T_2$ maps estimated using NNs trained with the typical mean squared error (MSE) criterion in comparison to the proposed method and dictionary matching. With only one noise realization ($N_r=1$), small $T_2$ values are poorly represented in the overall MSE loss, contributing to poor $T_2$ fits in vivo (E, consistent with Fig. 5 of Ref.~\citen{Cohen2018a}). While this is somewhat mitigated by averaging over $N_r=200$, the resulting $T_2$ maps are still biased (F), which is ameliorated by use of the proposed Bias-Reduced strategy (G).

\makeatletter \@input{xx.tex} \makeatother

\end{document}